\documentclass{osa-article}
\usepackage{siunitx}
\DeclareSIUnit\wn{\raiseto{-1}\cm}
\DeclareSIUnit\mum{\micro\m}
\newcommand{\ur}[1]{\text{#1}}
\journal{oe}
\begin{document}

\title{Fourier transform infrared spectroscopy with visible light}

\author{Chiara Lindner,\authormark{*} Sebastian Wolf, Jens Kie\ss ling, and Frank K\"{u}hnemann}

\address{Fraunhofer Institute for Physical Measurement Techniques IPM, Heidenhofstrasse 8, 79110 Freiburg, Germany\\}

\email{\authormark{*}chiara.lindner@ipm.fraunhofer.de} 

\begin{abstract}
Nonlinear interferometers allow spectroscopy in the mid-infrared range by detecting correlated visible light, for which non-cooled detectors with higher specific detectivity and lower dark count rates are available. We present a new approach for the registration of spectral information, which combines a nonlinear interferometer using non-degenerate spontaneous parametric down-conversion (SPDC) with a Fourier-transform spectroscopy concept. In order to increase the spectral coverage, we use broadband non-collinear SPDC in periodically poled LiNbO$_3$. Without the need for spectrally selective detection, continuous spectra with a spectral bandwidth of more than \SI{100}{\wn} are achieved. We demonstrate transmission spectra of a polypropylene sample measured with \SI{6}{\wn} resolution in the spectral range between \SIrange{3.2}{3.9}{\mum}.
\end{abstract}

\section{Introduction}
Mid-infrared spectroscopy is one of the most important techniques for chemical analysis of organic compounds. It finds a wide range of applications in industry and science. However, the detectors available for the mid-infrared range suffer from lower specific detectivities and higher dark count rates than detectors used for the visible or near-infrared range, and often require cooling \cite{Griffiths.2007}. There are several new concepts for pushing the boundaries of mid-infrared spectroscopy using nonlinear optical frequency conversion, such as upconversion spectroscopy \cite{Tidemand-Lichtenberg.2016,Wolf.2017} and difference-frequency-driven frequency-comb-based spectroscopic approaches \cite{Picque.2019,Kowligy.2019}.

Recently, nonlinear interferometers have sparked interest for their applications in quantum imaging \cite{Lemos.2014,Shih.2007}, optical coherence tomography \cite{Paterova.2018,Vanselow.2019} and spectroscopy\cite{Kalashnikov.2016b,Paterova.2017,Paterova.2018b}, allowing for measurements in the mid-infrared range by detecting entangled visible photons. They rely on spontaneous parametric down-conversion (SPDC), which can be described as the decay of a pump photon into two photons, called signal and idler, inside a nonlinear crystal. Signal and idler photons are correlated, and the sum of their frequencies is equal to the pump frequency. For an efficient emission, pump, signal and idler photons have to be phase matched. If the emission of two identical SPDC sources is overlapped, interference is observable in both, signal and idler beams.  This has been described by Zou, Wang and Mandel as induced coherence without induced emission \cite{Zou.1991} and is a direct consequence of the indistinguishability of the two photon pairs emitted by the identical sources. For two perfectly overlapped, coherently pumped SPDC sources with length $L$ and phase mismatch $\Delta k$, the intensity of the signal light can be described by\cite{Chekhova.2016}
\begin{equation}\label{eq:tpi}
I \propto \text{sinc}^2\left(\frac{\Delta k L}{2}\right)\left(1+\tau \cos(\Delta \varphi)\right).
\end{equation}
Therefore, the interference pattern depends on the transmission $\tau$ and phase difference $\Delta \varphi$ acquired by all three beams between the SPDC sources \cite{Burlakov.1997}.

Recently, the application of this effect for mid-infrared spectroscopy has been demonstrated in several publications \cite{Kalashnikov.2016b,Paterova.2017,Paterova.2018b}. Hereby, different interferometer geometries were used: A Mach-Zehnder set-up, consisting of two identical nonlinear crystals placed in a row \cite{Kalashnikov.2016b,Paterova.2017}, and a Michelson configuration, formed by a single nonlinear crystal with all beams passing the same crystal twice \cite{Paterova.2018b}. So far, two different methods for obtaining the spectral information of the mid-infrared idler from the correlated visible light have been demonstrated. As a first scheme, the signal light was analyzed with a spectrometer or spectrograph, which yielded a spectral resolution of \SI{20}{\wn}\cite{Kalashnikov.2016b} or \SI{5.2}{\wn}\cite{Paterova.2018b}. In the second method, the absorption coefficient and refractive index of a sample were determined from the change of the position and contrast of the interference fringes of the signal light detected with a silicon camera. The omission of the spectrometer, however, limited the spectral resolution of the measurement to the width of the interference fringes, which in that case was \SI{80}{\wn} \cite{Paterova.2017}.

In our work, we obtain the mid-infrared idler spectra via a Fourier transform of the visible signal interference pattern recorded while moving one interferometer mirror, allowing for simultaneous broadband measurements. The spectral bandwidth is increased by using non-collinear SPDC in periodically poled lithium niobate. Spectrally selective detection is not required, in analogy to classical Fourier-transform spectroscopy. Therefore, spectral resolution is limited only by the maximum optical path difference between the interferometer arms \cite{Griffiths.2007}. Additionally, \textit{a priori} knowledge is not required, neither of the absorption and refractive properties of the sample nor of the exact phase matching conditions of the SPDC process. In contrast to previous works, this allows measuring continuous spectra with both, a high bandwidth of more than \SI{100}{\wn} and a good resolution of \SI{6}{\wn} in a single setting.

\section{Methods}

\subsection{Setup}

\begin{figure}[t!]
	\centering\includegraphics[width=\linewidth]{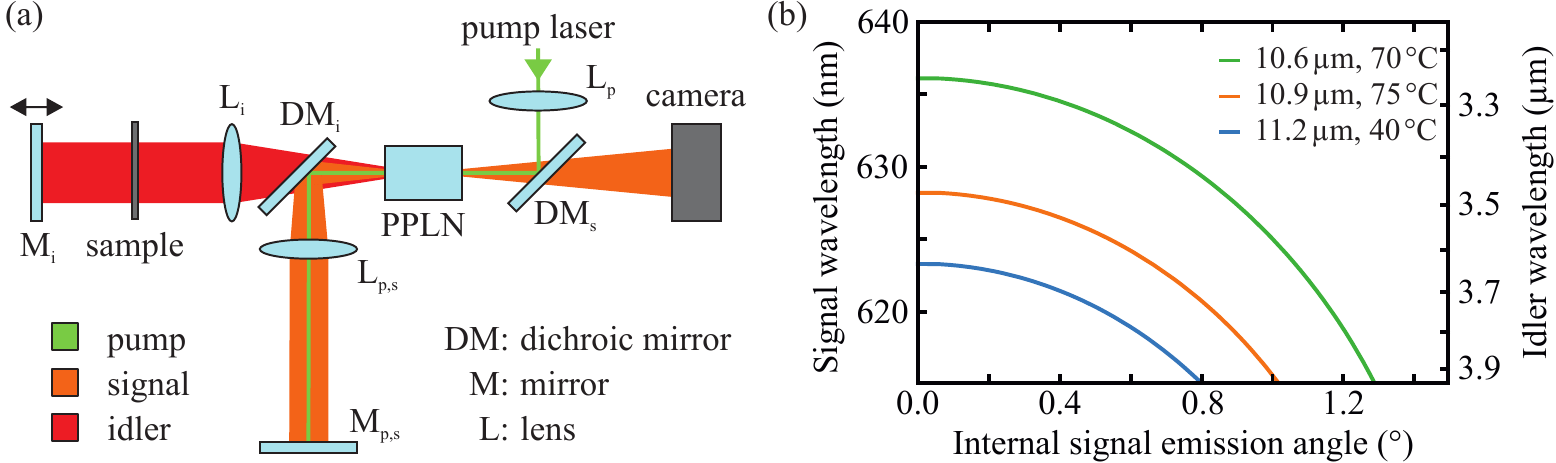}
	\caption{Nonlinear interferometer setup:\newline (a) Sketch of the setup. The pump laser (green line) is focused to the center of the nonlinear crystal (PPLN) and causes SPDC emission of the signal (orange) and idler (red) light. The beams are seperated using a dichroic mirror ($\ur{DM}_\ur{i}$), collimated ($\ur{L}_\ur{i}$ and $\ur{L}_\ur{p,s}$) and back-reflected by plane mirrors ($\ur{M}_\ur{i}$ and $\ur{M}_\ur{p,s}$). The camera records the signal interference pattern as a function of the displacement of the idler mirror. \newline (b) Angular dependent SPDC emission: Calculated phase matching curves (defined by $\Delta k=0$) for three different poling periods and crystal temperatures with a pump wavelength of \SI{532}{\nm} in lithium niobate. \label{fig:setup}}%
\end{figure}

A sketch of the experimental setup comprising a nonlinear interferometer in Michelson-geometry is shown in figure \ref{fig:setup}(a). As pump source we use a laser with \SI{532}{\nm} emission wavelength, a spectral linewidth of less than \SI{1}{\mega\Hz}, and up to \SI{2}{\W} output power (Coherent Verdi V2). The pump beam passes an optical isolator and is focused to a beam waist of \SI{120}{\mum} at the center of the nonlinear crystal. Reflected by the dichroic mirror $\ur{DM}_\ur{s}$, the pump passes through the nonlinear crystal, creating signal and idler photons via SPDC. As a nonlinear medium we use \SI{5}{\%} MgO-doped periodically poled lithium niobate (PPLN). The crystal is \SI{20}{\mm} long, \SI{0.5}{\mm} thick and contains 21 channels with different poling periods which are each \SI{0.5}{\mm} wide. The poling periods range from \SI{6.2}{\mum} to \SI{11.8}{\mum} in \SI{0.3}{\mum} steps. The crystal temperature is stabilized using a Peltier element.

Behind the crystal, the beams are split using the dichroic mirror $\ur{DM}_\ur{i}$ which reflects visible and near-infrared light and transmits mid-infrared light. The transmitted idler beam is collimated with a $\ur{CaF}_2$ lens ($\ur{L}_\ur{i}$) with \SI{50}{\mm} focal length. The plane mirror $\ur{M}_\ur{i}$ placed in \SI{50}{\mm} distance to the lens reflects the idler light, effectively forming a 4$f$ relay optic imaging back into the crystal center. The end mirror is mounted on a piezo positioning system with a maximum displacement of \SI{800}{\mum} in beam direction and a position resolution of \SI{1.8}{\nm}. A spectroscopic sample can be placed in the collimated idler light in front of the plane mirror. The signal and pump beams are reflected on the dichroic mirror $\ur{DM}_\ur{i}$ into a second 4$f$ optical path of equal length, where they are collimated by a lens $\ur{L}_\ur{p,s}$ and reflected by a (fixed) plane mirror $\ur{M}_\ur{p,s}$.

The back-travelling pump beam causes a second SPDC process, which is indistinguishable from the first SPDC emission if the beams are correctly aligned. We verified experimentally that the superposition with the first SPDC emission has no effect on the total rates of the second SPDC process by comparing the total count rates on the camera sensor with and without blocking the idler path, meaning that no stimulated emission occurs. Behind the crystal, the pump light is again reflected by the dichroic mirror $\ur{DM}_\ur{s}$ and then removed by the optical isolator to prevent damage to the pump laser source. The signal beam is transmitted by the dichroic mirror $\ur{DM}_\ur{s}$ and passes spectral filters, which remove residual pump, idler, and ambient light. The interference pattern of the signal light is then detected by a silicon sCMOS camera (Andor Zyla 4.2).

\subsection{Phase matching considerations}

The SPDC emission is governed by the phase mismatch $\Delta k$ between the pump, signal and idler waves. Efficient emission is only possible with small phase mismatch, which can be controlled by the crystal poling period and temperature. For a poling period of $\Lambda$, the phase mismatch can be calculated by its longitudinal ($z$) and vertical ($x$) components (with respect to the pump beam direction):
\begin{eqnarray}
\Delta k_z &=& k_\ur{p} - 2\pi/\Lambda -k_\ur{s} \cos(\theta_\ur{s}) - k_\ur{i} \cos(\theta_\ur{i}) \nonumber \\
\Delta k_x &=& k_\ur{s} \sin(\theta_\ur{s}) - k_\ur{i} \sin(\theta_\ur{i}) \newline \nonumber \\
\Delta k &=& \sqrt{\Delta k_x^2+\Delta k_z^2}
\end{eqnarray}
Hereby, $k_\ur{p,s,i}=2\pi n_\ur{p,s,i}/\lambda_\ur{p,s,i}$ denotes the length of the wave vector for refractive index $n_\ur{p,s,i}$ and wavelength $\lambda_\ur{p,s,i}$ of pump, signal and idler light, respectively. The internal emission angles $\theta_\ur{s,i}$ for signal and idler beams are defined with respect to the pump beam direction.  

Figure \ref{fig:setup}(b) shows calculated curves of phase-matched ($\Delta k=0$) signal emission wavelengths depending on the internal signal emission angle $\theta_\ur{s}$ for three different poling periods and crystal temperatures. A signal wavelength range of \SIrange{615}{640}{\nm} with a pump wavelength of \SI{532}{\nm} corresponds to idler wavelengths ranging from \textasciitilde ~\SIrange{3.9}{3.2}{\mum}, which is an interesting spectral range for the analysis of organic compounds, due to the strong C-H fundamental vibration bands. The phase matching curves show that by imaging over a large emission angle range, one is able to obtain a broad spectral bandwidth. It is to be noted, that the idler emission angle $\theta_\ur{i}$ is bigger than the signal emission angle $\theta_\ur{s}$, since $\theta_\ur{i}\approx \lambda_\ur{i}/\lambda_\ur{s}\cdot \theta_\ur{s}$. Both external emission angles are further increased due to refraction on the crystal facet.  
\subsection{Measurement procedure and analysis}
\begin{figure}[ht]
	\centering\includegraphics[width=\linewidth]{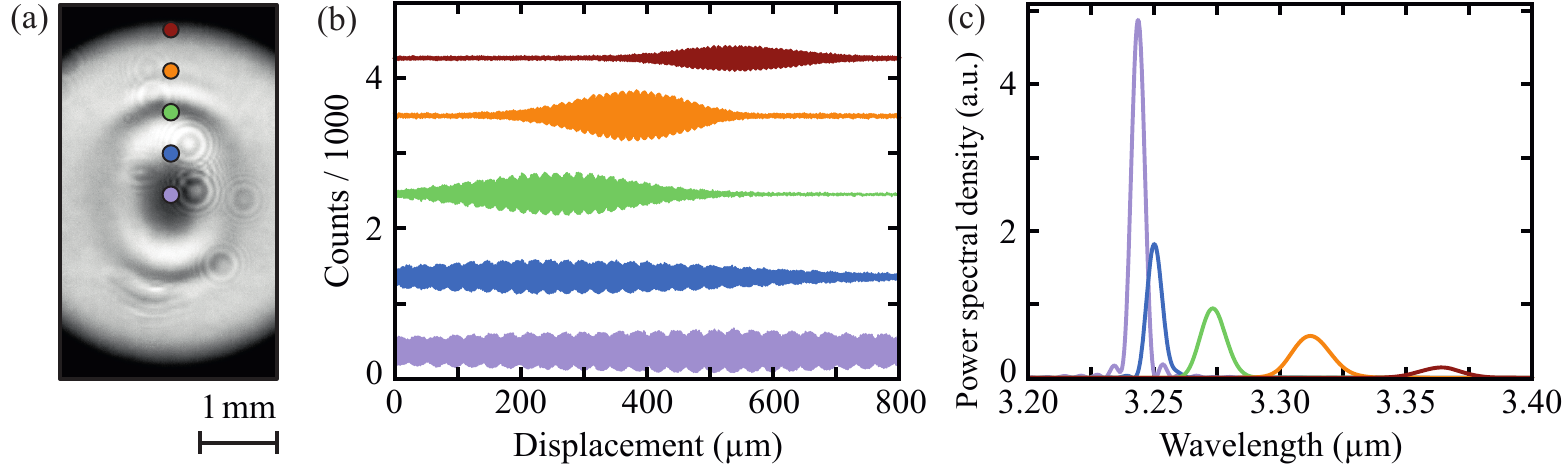}%
	\caption{Signal interference pattern, single-pixel interferograms, and Fourier transformed spectra of the reference measurement with \SI{10.6}{\mum} poling period and \SI{70}{\degreeCelsius} crystal temperature: \newline
		(a) Single camera image of the signal interference pattern at \SI{400}{\mum} displacement, rotated counterclockwise by \SI{90}{\degree}. Indicated are five pixel positions, for which the interferograms and spectra are shown in (b) and (c).\newline
		(b) Interferograms for the selected camera pixels as indicated in (a), recorded for a total idler mirror displacement of \SI{800}{\mum}. The interferograms are shifted vertically for clarity in 1000-count-steps.  The superstructure oscillating with low frequencies is due to aliasing effects.\newline 
		(c) Spectra (power spectral density) of the interferograms shown in (b), obtained by discrete Fourier transform. 
		\label{fig:coherence}}
\end{figure}

The poling period and crystal temperature are set for efficient near-collinear emission at the targeted idler wavelength range. The idler mirror is moved in beam direction and the resulting change in path length difference between the signal and idler interferometer arms is determined using the position reference of the piezo positioning system. A camera image (400 x 800 pixel) of the signal interference pattern is recorded with \SI{20}{\ms} integration time for each \SI{400}{\nm} step of displacement. Up to the maximum displacement of \SI{800}{\mum} (equivalent to an optical path difference of \SI{1.6}{\mm}) 2000 camera images are taken, resulting in a total data acquisition time of \SI{40}{\s}.

The interference pattern of a measurement taken with a poling period of \SI{10.6}{\mum} and a crystal temperature of \SI{70}{\degreeCelsius} is shown in Fig. \ref{fig:coherence}(a). The interference fringes exhibit a near circular pattern. We attribute the missing interference contrast in vertical direction to the narrow crystal aperture in this direction.

Figure \ref{fig:coherence}(b) shows the detected intensity at five different positions on the camera sensor (marked with dots of the corresponding color in Fig. \ref{fig:coherence}(a)) as a function of the idler mirror displacement. The detected signal light is modulated with the idler frequency as the idler mirror is displaced, varying the phase (see Eq. \ref{eq:tpi}). Two effects can be noticed: At near-collinear emission (violet curve), the interferogram shows high modulation contrast along the whole displacement range. For increasing distance to the center (i.e. larger emission angles), the envelope of the interferogram becomes narrower. This indicates a larger spectral bandwidth of the light detected by the respective pixel, which reduces the coherence length. The reason behind this can be seen in the phase matching curve shown in Fig. \ref{fig:setup}(b): With an increasing emission angle, the phase matching curve becomes steeper, thus a small angle element (as the one measured by one camera pixel) includes a larger wavelength range.
Secondly, with increasing emission angle, the point with highest interference contrast changes its position along the displacement axis. This is caused by a changing path difference between signal and idler beam for increasing emission angles, due to dispersion inside the nonlinear crystal and geometric effects. 

In our measurement approach, the spectral information (wavelength and intensity) is obtained purely by Fourier transformation of the interferograms recorded when moving the mirror M$_\ur{i}$. For this, partial spectra are calculated from the interferogram of each camera pixel by a discrete Fourier Transform (DFT) based on the Fast Fourier Transform (FFT) algorithm. This is demonstrated on the interferograms from Fig. \ref{fig:coherence}(b), the Fourier transformed spectra are plotted in Fig. \ref{fig:coherence}(c) in the same color. With increasing distance from the center (violet to dark red), the partial spectra shift to longer wavelengths. This is in agreement with the non-collinear phase matching conditions (Fig. \ref{fig:setup}(b)). The increasing width of the partial spectra, which is also visible in the interferograms, is also a consequence of phase matching. For the near-collinear interferogram (violet), side bands are visible in the Fourier transformed spectrum. These are caused by the finite displacement range of the interferogram, which results in a rectangular window function. Each of the individual Fourier-transformed spectra has a peak signal-to-noise ratio of more than $10^4$ at a total measurement time of \SI{40}{\s} (2000 frames each taken with \SI{20}{\ms} integration time per frame).
It is important to note, that even though the wavelength and bandwidth of the emission changes with the emission angle, the performance of the spectral analysis does not - in contrast to many upconversion spectroscopy schemes using non-collinear phase matching \cite{Tidemand-Lichtenberg.2016, Wolf.2017}.

\begin{figure}[t!]
	\centering\includegraphics[width=\linewidth]{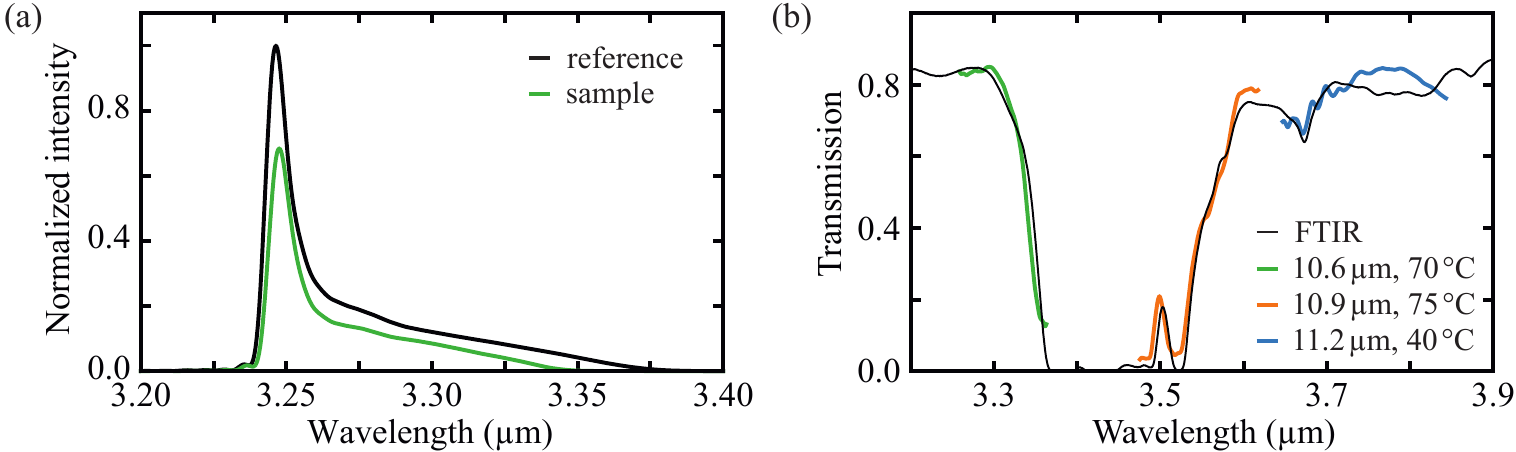}%
	\caption{Measurement results: \newline
		(a) Reference and sample spectrum measured with \SI{10.6}{\mum} poling period and \SI{70}{\degreeCelsius} crystal temperature. \newline
		(b) Transmission spectrum of a thin polypropylene film. Black line: Measurement taken with commercial Fourier-transform infrared spectrometer. Colored lines: Transmission calculated from the spectra measured with the nonlinear interferometer with different quasi phase matching conditions. \label{fig:results}}
\end{figure}
To obtain the complete spectrum, first, the Fourier transforms are calculated for each pixel individually, and in the next step, the Fourier transforms of all camera pixels are summed.
This pixel-wise approach is necessary for the following reason: As one can see in Fig. \ref{fig:coherence}(b), the maxima of the interferograms for the different pixels are observed at different values of idler mirror displacement. This points towards a radially varying phase and optical path in the present setup. If one were to add up the interferograms of all camera pixels (or focus the light onto a single-pixel detector) and perform a single Fourier transformation afterwards, these phase variations would erase the spectral information. 
As a consequence, the spectral information contained in the individual pixels needs to be extracted first via the Fourier transformation before the total spectrum can be obtained via summation over all partial spectra. 
The black line in Fig. \ref{fig:results}(a) shows the total spectrum calculated from the sum of the Fourier transform of all 400 x 800 camera pixels in the interference pattern shown in Fig. \ref{fig:coherence}(a). The spectrum has a peak in intensity at \SI{3.25}{\mum} wavelength, which is the phase matched collinear idler wavelength (see Fig. \ref{fig:setup}(b)). For larger wavelengths, the spectral intensity decreases. As can be seen from Fig. \ref{fig:coherence}(c), longer wavelengths are observed for larger emission angles, which are prone to imaging errors and reduced mode overlap, resulting in a smaller interference contrast. 

To determine the transmission spectrum of a sample, the measurement procedure described above is then repeated with a sample introduced into the collimated idler beam. For demonstration purposes we chose a polypropylene film (thickness $\approx$\SI{120}{\mum}), which has an absorption band around \SI{3.4}{\mum}. The sample spectrum is shown in Fig. \ref{fig:results}(a) as a green line. For the whole wavelength range, a reduced intensity of the sample spectrum is visible. We restrict both reference and sample spectrum to the non-collinear spectral range where no side bands of the Fourier-transformed spectra occur. Since the measurement sample is passed twice inside the interferometer, the transmission $T$ is then calculated from the reference $I_\ur{r}$ and sample $I_\ur{s}$ spectrum by:
\begin{equation}
T= \sqrt{\frac{I_\ur{s}}{I_\ur{r}}}
\end{equation}
An intensity threshold in the reference spectrum at \SI{2}{\percent} of the maximum intensity was chosen to limit the evaluated spectral ranges for transmission measurements. These measurements are repeated for the three phase matching settings shown in Fig. \ref{fig:setup}(b). For qualitative and quantitative comparison, the transmission of the polypropylene sample is also measured in a Fourier-transform infrared (FTIR) spectrometer (Bruker Vertex 80) with 64 scans at a resolution of \SI{0.5}{\wn}.

\section{Results}

Figure \ref{fig:results}(b) shows the transmission spectra of the thin polypropylene film. The black line represents the measurement of the sample taken with the commercial FTIR spectrometer (Bruker Vertex 80). The transmission spectra measured with the nonlinear interferometer at three different phase matching settings are shown as colored lines (poling periods and crystal temperatures according to the color code). The spectra show good agreement to the FTIR measurement, they are continuous and have a spectral coverage of more than \SI{100}{\wn} each. Due to the high signal-to-noise ratio of the spectra (more than $10^4$ at a total integration time of \SI{40}{\s}), the sensitivity of the measurement approach is possibly not limited by noise or signal strength, but rater by systematic deviations and instabilities of the interferometer itself. Since the spectral information in our measurement approach is obtained from a spatial-domain Fourier transformation, the spectral resolution is determined solely by the idler mirror displacement. In our setup, the spectral resolution is limited to $\approx$ \SI{6}{\wn}, which is comparable to the resolution demonstrated in Ref. \cite{Paterova.2018b}. The spectral resolution can be increased by using a different type of translation stage allowing larger mirror displacement.

\section{Conclusion}

In a nonlinear interferometer, the amplitude or phase information acquired by the idler photons can be measured by detecting the signal photons. With an interferometer in Michelson geometry, and a measurement scheme closely related to classical Fourier-transform infrared spectroscopy, it is possible to record the spectral information without the use of spectral selection or beforehand know\-ledge of the signal and idler wavelengths. As shown in the experiments, the measured transmission spectrum of a polymer sample is in good agreement with a conventionally measured reference spectrum. Utilizing a large wavelength difference between signal and idler photons, the infrared information is measured by detecting the correlated visible light of spontaneous parametric down-conversion. This approach allows one to perform infrared spectroscopy using a visible laser light source and a silicon-based detector. The latter circumvents the usage of expensive infrared detectors, which often have to be thermoelectrically or even cryogenically cooled.
The demonstrated measurement technique can easily be extended to the whole infrared transparency range of lithium niobate; possibly even into the terahertz regime \cite{Haase.2019,Kutas.2019}. The use of other nonlinear materials allows the extension of the technique to the far infrared. It is expected that the presented approach, comprising the combination of nonlinear interferometers and Fourier transform analysis, will provide easily applicable spectroscopy methods for wide fields in industry and science.

\section*{Funding}
This work was supported through the Fraunhofer Lighthouse Project QUILT.

\section*{Disclosures}
The authors declare no conflicts of interest.

\bibliography{literature}

\begin{thebibliography}{10}
\newcommand{\enquote}[1]{``#1''}

\bibitem{Griffiths.2007}
P.~R. Griffiths and J.~A. de~Haseth, \emph{Fourier transform infrared
  spectrometry} (Wiley-Interscience, 2007).

\bibitem{Tidemand-Lichtenberg.2016}
P.~Tidemand-Lichtenberg, J.~S. Dam, H.~V. Andersen, L.~H{\o}gstedt, and
  C.~Pedersen, \enquote{Mid-infrared upconversion spectroscopy,}
  {\protect\JournalTitle{J. Opt. Soc. Am. B}} \textbf{33}, D28--D35 (2016).

\bibitem{Wolf.2017}
S.~Wolf, J.~Kiessling, M.~Kunz, G.~Popko, K.~Buse, and F.~K\"{u}hnemann,
  \enquote{Upconversion-enabled array spectrometer for the mid-infrared,
  featuring kilohertz spectra acquisition rates,} {\protect\JournalTitle{Opt.
  Express}} \textbf{25}, 14504--14515 (2017).

\bibitem{Picque.2019}
N.~Picqu\'{e} and T.~W. H\"{a}nsch, \enquote{Frequency comb spectroscopy,}
  {\protect\JournalTitle{Nature Photonics}} \textbf{13}, 146--157 (2019).

\bibitem{Kowligy.2019}
A.~S. Kowligy, H.~Timmers, A.~J. Lind, U.~Elu, F.~C. Cruz, P.~G. Schunemann,
  J.~Biegert, and S.~A. Diddams, \enquote{Infrared electric field sampled
  frequency comb spectroscopy,} {\protect\JournalTitle{Science Advances}}
  \textbf{5} (2019).

\bibitem{Lemos.2014}
G.~B. Lemos, V.~Borish, G.~D. Cole, S.~Ramelow, R.~Lapkiewicz, and
  A.~Zeilinger, \enquote{Quantum imaging with undetected photons,}
  {\protect\JournalTitle{Nature}} \textbf{512}, 409--412 (2014).

\bibitem{Shih.2007}
Y.~Shih, \enquote{Quantum imaging,} {\protect\JournalTitle{IEEE Journal of
  Selected Topics in Quantum Electronics}} \textbf{13}, 1016--1030 (2007).

\bibitem{Paterova.2018}
A.~V. Paterova, H.~Yang, C.~An, D.~A. Kalashnikov, and L.~A. Krivitsky,
  \enquote{Tunable optical coherence tomography in the infrared range using
  visible photons,} {\protect\JournalTitle{Quantum Science and Technology}}
  \textbf{3}, 025008 (2018).

\bibitem{Vanselow.2019}
A.~Vanselow, P.~Kaufmann, I.~Zorin, B.~Heise, H.~Chrzanowski, and S.~Ramelow,
  \enquote{Mid-infrared frequency-domain optical coherence tomography with
  undetected photons,} in \emph{Quantum Information and Measurement (QIM) V:
  Quantum Technologies,}  (Optical Society of America, 2019), p. T5A.86.

\bibitem{Kalashnikov.2016b}
D.~A. Kalashnikov, A.~V. Paterova, S.~P. Kulik, and L.~A. Krivitsky,
  \enquote{Infrared spectroscopy with visible light,}
  {\protect\JournalTitle{Nature Photonics}} \textbf{10}, 98--101 (2016).

\bibitem{Paterova.2017}
A.~Paterova, S.~Lung, D.~A. Kalashnikov, and L.~A. Krivitsky,
  \enquote{Nonlinear infrared spectroscopy free from spectral selection,}
  {\protect\JournalTitle{Scientific reports}} \textbf{7}, 42608 (2017).

\bibitem{Paterova.2018b}
A.~Paterova, H.~Yang, C.~An, D.~Kalashnikov, and L.~Krivitsky,
  \enquote{Measurement of infrared optical constants with visible photons,}
  {\protect\JournalTitle{New Journal of Physics}} \textbf{20}, 043015 (2018).

\bibitem{Zou.1991}
X.~Y. Zou, L.~J. Wang, and L.~Mandel, \enquote{Induced coherence and
  indistinguishability in optical interference,}
  {\protect\JournalTitle{Physical review letters}} \textbf{67}, 318--321
  (1991).

\bibitem{Chekhova.2016}
M.~V. Chekhova and Z.~Y. Ou, \enquote{Nonlinear interferometers in quantum
  optics,} {\protect\JournalTitle{Advances in Optics and Photonics}}
  \textbf{8}, 104 (2016).

\bibitem{Burlakov.1997}
A.~V. Burlakov, M.~V. Chekhova, D.~N. Klyshko, S.~P. Kulik, A.~N. Penin, Y.~H.
  Shih, and D.~V. Strekalov, \enquote{Interference effects in spontaneous
  two-photon parametric scattering from two macroscopic regions,}
  {\protect\JournalTitle{Physical Review A}} \textbf{56}, 3214--3225 (1997).

\bibitem{Haase.2019}
B.~Haase, M.~Kutas, F.~Riexinger, P.~Bickert, A.~Keil, D.~Molter, M.~Bortz, and
  G.~von Freymann, \enquote{Spontaneous parametric down-conversion of photons
  at 660 nm to the terahertz and sub-terahertz frequency range,}
  {\protect\JournalTitle{Optics express}} \textbf{27}, 7458--7468 (2019).

\bibitem{Kutas.2019}
M.~Kutas, B.~Haase, P.~Bickert, F.~Riexinger, D.~Molter, and G.~von Freymann,
  \enquote{Terahertz quantum sensing,}
  {\protect\JournalTitle{arXiv:1909.06855}}  (2019).

\end{thebibliography}

\end{document}